\documentclass[aip,amsmath,amssymb,reprint, jcp]{revtex4-1}
\usepackage{graphicx}
\usepackage{dcolumn}
\usepackage{bm}
\usepackage{multirow}					
\usepackage{array}
\usepackage{footnote} 

\usepackage{booktabs}
\usepackage{float}
\usepackage{mathtools}
\usepackage{color}
\usepackage{natmove}
\usepackage{lipsum}
\DeclareMathOperator*{\argmin}{\arg\!\min}
\usepackage{listings}
\usepackage{pythonhighlight}
\usepackage{booktabs}
\usepackage{xcolor}
\usepackage{subfig}
\definecolor{codegreen}{rgb}{0,0.6,0}
\definecolor{codegray}{rgb}{0.5,0.5,0.5}
\definecolor{codepurple}{rgb}{0.58,0,0.82}
\definecolor{backcolour}{rgb}{0.95,0.95,0.92}
\lstdefinestyle{mystyle}{
    backgroundcolor=\color{backcolour},   
    commentstyle=\color{codegreen},
    keywordstyle=\color{magenta},
    numberstyle=\tiny\color{codegray},
    stringstyle=\color{codepurple},
    basicstyle=\ttfamily\footnotesize,
    breakatwhitespace=false,         
    breaklines=true,                 
    captionpos=b,                    
    keepspaces=true,                 
    numbers=left,                    
    numbersep=5pt,                  
    showspaces=false,                
    showstringspaces=false,
    showtabs=false,                  
    tabsize=2
}
\begin{document}
\newcommand{\agoxrepo}{\url{https://gitlab.com/agox/agox}}
\newcommand{\agoxdocu}{\url{https://agox.gitlab.io/agox}}
\newcommand{\agoxdata}{\url{https://gitlab.com/agox/agox_data}}
\newcommand{\license}{GNU GPLv3}

\title[]{Atomistic Global Optimization X: A Python package for optimization of atomistic structures}
\author{Mads-Peter V. Christiansen}
\author{Nikolaj Rønne}
\author{Bj{\o}rk Hammer}
\email{hammer@phys.au.dk}
\affiliation{Center for Interstellar Catalysis, Department of Physics and Astronomy, Aarhus University, DK-8000 Aarhus, Denmark}
\begin{abstract}
Modelling and understanding properties of materials from first
principles require knowledge of the underlying atomistic
structure. This entails knowing the individual chemical identity and position
of all atoms involved. Obtaining such information for
macro-molecules, nano-particles, clusters, and for the surface,
interface, and bulk phases of amorphous and solid materials
represents a difficult high-dimensional global optimization
problem. The rise of machine learning techniques in materials
science has, however, led to many compelling developments that may
speed up structure searches. The complexity of such new methods
has prompted a need for an efficient way of assembling them into global optimization
algorithms that can be experimented with. In this paper, we introduce the Atomistic Global
Optimization X (AGOX)
framework and code, as a customizable approach that enables efficient
building and testing of global optimization algorithms. A modular way of expressing global 
optimization algorithms is described and modern programming practices are used to enable that modularity in the freely 
available AGOX python package. A number of examples of global optimization
approaches are implemented and analyzed. This ranges from random
search and basin-hopping to machine learning aided approaches with
on-the-fly learnt surrogate energy landscapes. The methods are
show-cased on problems ranging from supported clusters over surface
reconstructions to large carbon clusters and metal-nitride clusters
incorporated into graphene sheets.
\end{abstract}
    
\maketitle

\section{Introduction}
Global optimization is a prerequisite for the computational treatment of materials at the atomic level. The application 
of quantum mechanics to predict the properties of materials requires
knowledge of the positions of all the atoms that make up the material. 
The positions that occur in reality will often be very close to those that minimize the total quantum mechanical energy, whereby the 
importance of global optimization for atomistic structure is evident. The problem may be visualized as finding the lowest 
valley in a high-dimensional landscape, the potential energy surface (PES). This surface may be described at different levels of theory, from 
crude distance-based pair-potentials to sophisticated calculations treating the particles at the quantum level. With increased accuracy 
comes increased computational cost, therefore global optimization methods that efficiently search the PES for the lowest energy minima are necessary. 
Computational approaches that utilize global optimization have received significant attention and led to impressive results \cite{Greeley2006, Pickard2006,Piazza2014,Jain2016,Oganov2019}.

A multitude of optimization algorithms have been proposed from simulated annealing \cite{Kirkpatrick1983} through basin-hopping \cite{Wales1997} and minima-hopping \cite{Goedecker2004} methods to evolutionary 
algorithms \cite{Deaven1995,Johnston2003, Oganov2006, Wu2013, Vilhelmsen2014, risko} to ab-initio-random structure search \cite{Pickard2011} and
particle swarm algorithms \cite{Calypso1,Calypso2} to mention a few of the most successful methods. In recent years, machine learning has become a central 
topic in computational materials science. A prominent example of which is the ever increasing accuracy of machine learned potential 
energy surfaces, colloquially referred to simply as machine learning potentials \cite{Behler2007, Bartok2010, Behler2011, Rupp2012, Bartok2013, Hansen2015, Faber2017, Smith2017, Deringer1, Schutt2018, Lubbers2018, Deringer2, Lei2020, Zaverkin2021,Timmermann2021, hu2021}. 
Machine learning has also led to substantial improvements for simulation tasks in computational material science,  
such as machine learning potential based molecular-dynamics simulations for investigating properties of materials \cite{Kermode2015, Gastegger2017, Deringer2018_A, Deringer2018_B, Jinnouchi2019_A, Noe2020, Lim2020, Boselt2021}, 
optimization algorithms exploiting machine learning potentials \cite{Runhai2015, Patra2016, NNsearch:alexandrova, Jindal2017, Deringer2018, Denzel2018, Rio2018, Schmitz2018, Tong2018, Kolsbjerg2018, Jacobsen2018Local, Milica2019, Jennings2019, Evgeny2019, Behler2020, Bisbo2020, BEACON, Arrigoni2021, Yilin2021, Vaidish2021, goldsmith2022}
or machine learning methods for guided exploration of the PES \cite{Jorgensen2017Clustering, Meldgaard2018, Jorgensen2018Bayes, Sorensen2018, Jorgensen2018ASLA, Pickard2019, Chiriki2019, Zhou2019, Meldgaard2020, Simms2020_1, Simms2020_2,BEACON_INTER}.

With the speed of advancement in the materials science and machine
learning communities it is essential that software tools are available
that allow quick experimentation. This is especially true for global
optimization (GO), an open-ended subject with room for new and
improved algorithms. In GO, the aim is to identify the optimum
solution for a complex target function. Being goal-driven, any GO
method that does so is legitimate if the goal is eventually reached,
and hence experimentation with the computational strategy is
welcomed. This contrasts other tasks in materials science, such as
solving the Kohn-Sham equations and calculating forces in density
functional theory calculations, or propagating atomic positions in
molecular dynamics simulations, where the methods are governed by
well-established defining equations, and where efficient algorithms and
codes have been developed.

\begin{figure*}
    \centering
    \includegraphics[trim={0cm 0cm 0 0},clip]{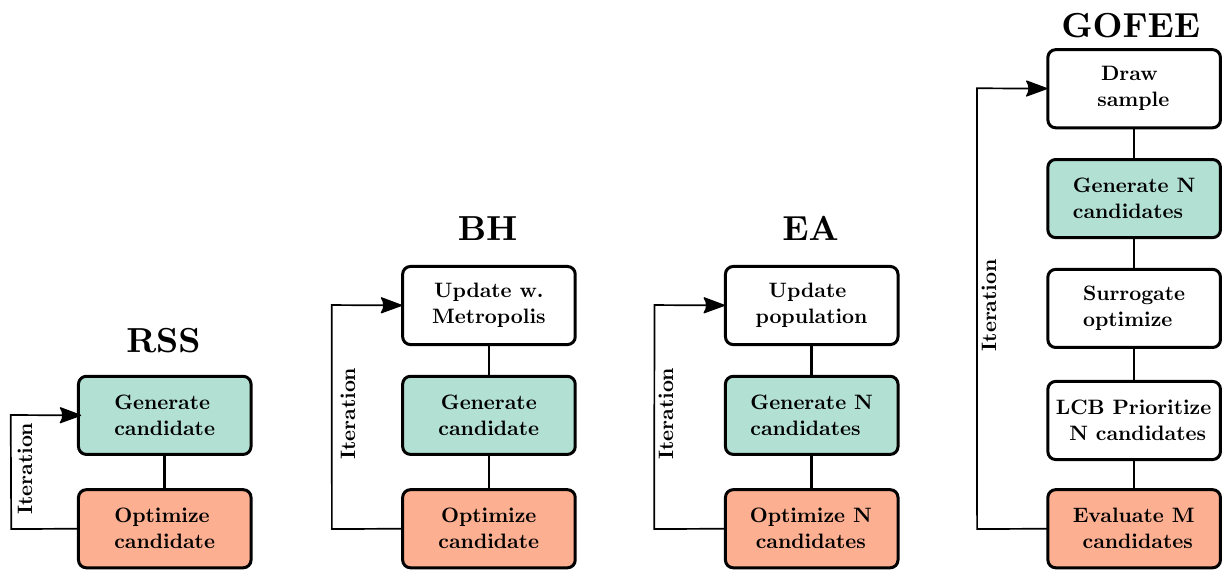}
    \caption{Overview of several popular global optimization algorithms. Random structure search (RSS) represents 
    the simplest method. In each iteration of RSS, a structural
    candidate is generated and locally optimized.
    In basin-hopping (BH), the current position in the search-space is kept track of and 
    updated using the Metropolis criterion. The position is then used by the generation mechanism. 
    In the evolutionary algorithm (EA), a population of candidates is maintained of and serves as input to the
    generation of a new candidate in each iteration which is then optimized.
    In GOFEE, a number of candidates are generated 
    and are locally optimized in a computationally inexpensive surrogate potential before 
    deciding on which candidate to evaluate in the true potential using a 
    lower confidence bound acquisition function. The orange boxes will for 
    many problems involve computationally expensive potentials such as DFT.}
    \label{fig:agox_flowchart}
\end{figure*}

In this work, we introduce the Atomistic Global Optimization X (AGOX) framework and python package. AGOX is modular 
and flexible such that many popular global optimization algorithms can be formulated in the framework and realized 
in the code. The package builds on the atomistic simulation environment \cite{Larsen2017} (ASE) and can thus be used with a multitude 
of electronic structure codes, with a similar focus on effortless scripting as ASE enables. An overview of the 
framework is presented, and the goals of the code are discussed in detail. We present applications of the code on several 
systems, starting with a cluster on a metal surface described by a cheap potential that is solved using four different 
global optimization algorithms. A second application is used to highlight the ability of AGOX to surgically make 
algorithmic changes for a tin-oxide surface system, where basin-hopping and a machine learning enhanced basin-hopping 
algorithm are compared. The third example is used to discuss parallelization options, where parallel tempering \cite{Kofke2002} is used to 
solve a two-dimensional carbon cluster show casing how it is possible to take advantage of computational resources to 
reduce waiting time for results. The fourth and final example documents considerations taken when solving a computationally
demanding problem, in this case a metal-nitride cluster embedded in a
graphene sheet where also spin polarization must be taken into account.
The AGOX code is freely available on gitlab \agoxrepo.

\section{Method}
\subsection{AGOX Framework}
\label{sec:AGOX_framework}

A large number of global optimization methods have been proposed, some of which differ only slightly
while others differ significantly, however, all of them involve
two essential steps.
The first step is the generation of a candidate structure and the second is the evaluation 
of the generated candidate in the target potential. In Figure
\ref{fig:agox_flowchart} examples of global optimization methods
having these two steps are given. The first example given is random
structure search (RSS),
that consists of just candidate generation and local optimization. The
next examples are basin-hopping (BH) and evolutionary algorithms (EA)
that both use previous structures as the starting point for the generation, i.e. a perturbation of the atomic positions or as a crossover mutation that combines two or more structures. They therefore need to keep 
track of which previous structures are used in this way. The final
example is the global
optimization with first-principles energy expressions (GOFEE)
method\cite{Bisbo2020}, in whih the expensive local optimization in the 
target potential is replaced by local optimization in an on-the-fly trained surrogate potential. Due to 
the much reduced cost, several candidates may be optimized per iteration and a small number of them may be selected 
for evaluation in the target potential. 

\begin{figure*}
    \includegraphics[width=0.9\textwidth]{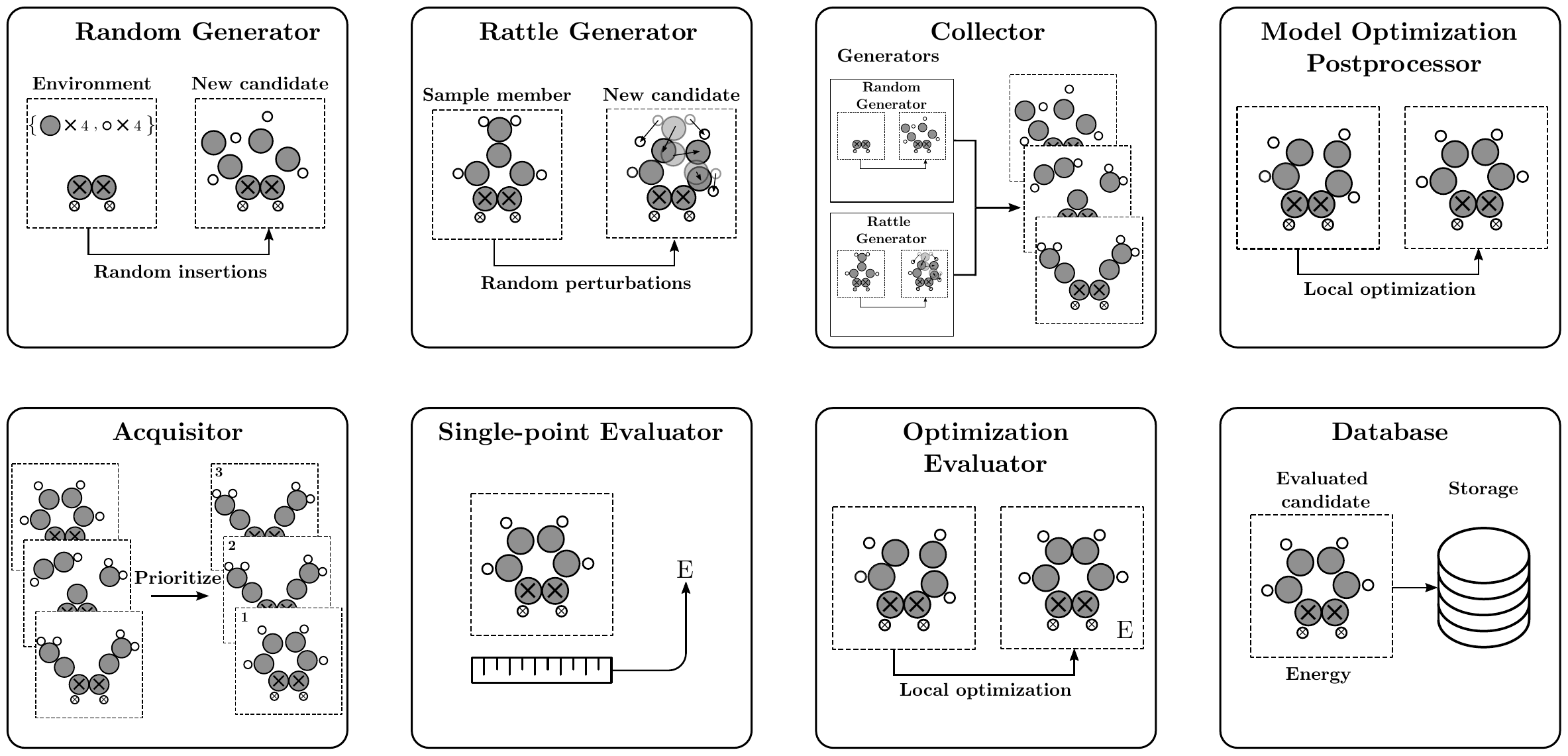}
    \caption{Pictorial illustration some of the action-type AGOX modules.
    \protect\fbox{\sc {generators}} generate candidate structures, a random generator places atoms at random with the only 
    requirement that bond lengths are within are not too short or too long, whereas a rattle generator perturbs a previously evaluated structure. 
    A \protect\fbox{\sc {collector}} may be used to collect several 
    generated candidates. \protect\fbox{\sc {postprocessors}} may be used to improve generated candidates regardless of which 
    generator they originate from, such as local optimization in a model or moving to the center of a cell. 
    An \protect\fbox{\sc {acquisitor}} can be used to select the most promising candidate. \protect\fbox{\sc {evaluators}} calculate key properties of 
    candidates, e.g. the energy and the evaluated structure is stored in a \protect\fbox{\sc {database}}.}
\end{figure*}

AGOX is a framework that allows all of these algorithms, and more, to be used within a single codebase. This is achieved by defining a 
number of modules that can be used to build this wide range of
optimization algorithms. At present, we have identified the need for
two data-type modules and eight action-type modules. The data-type
modules enable the handling of the candidates and are as follows:

The \fbox{\sc {environment}} module handles the simulation cell, any already present atoms (which we call a template) and the number and species of 
the atoms the search algorithm should place. This is the fundamental
module that defines the properties of the global optimization problem.

The \fbox{\sc {candidate}} module manages all the information about the
structural candidates, e.g. position of the atoms by inheriting functionality from the 
ASE Atoms object, and by what means the candidate originated. 

The action-type modules perform actions based on the structural
information in objects from the data-type modules and do in some cases update
one or several candidate objects. The eight action-type objects are:

The \fbox{\sc {database}} module stores the proposed solutions, that obey
the conditions defined by the environment. These may be candidates
whose properties have not yet been calculated in the target potential, or
structures for which such properties are indeed available. The database represents all the
knowledge gathered about the target potential that can be analyzed at
the end of the search to determine the global minimum energy structure.
Furthermore, several other modules can leverage the information stored
in the database during the search, e.g. to build machine learning
potentials, to maintain a population, or to extract a sample. 

The \fbox{\sc {model}} module builds machine learning potentials, such as the gaussian process regression model employed by GOFEE, based on the structures stored in the database.

\begin{figure}
    \includegraphics[width=0.48\textwidth]{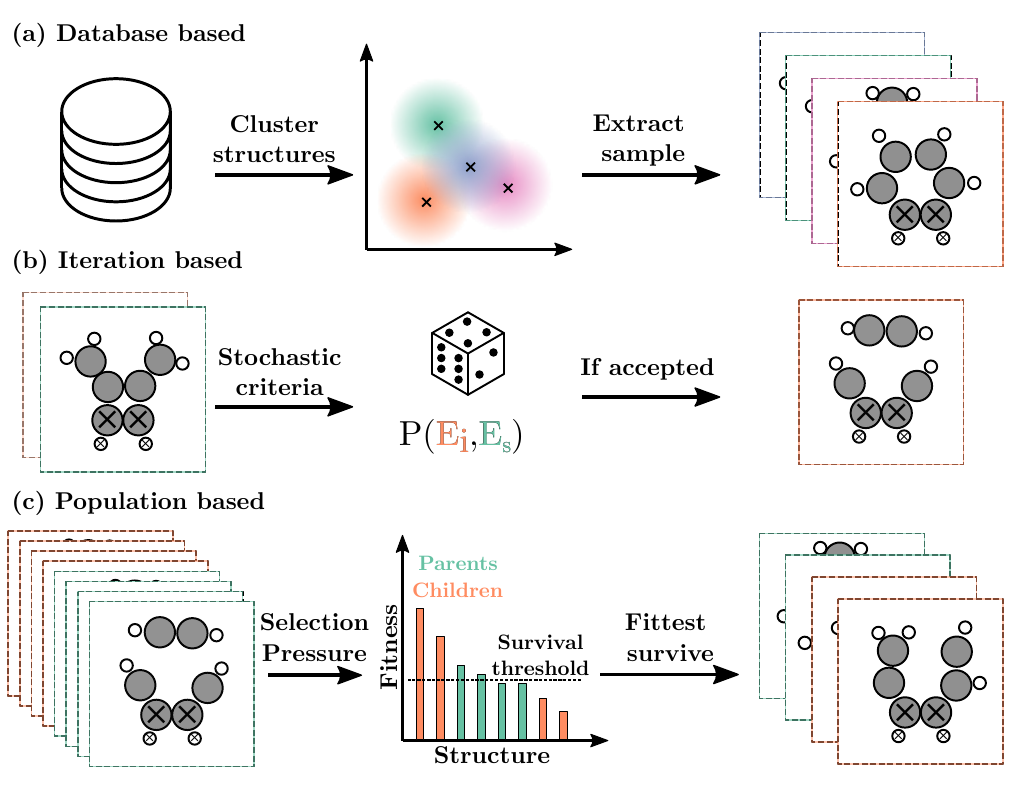}
    \caption{The starting point of candidate generation can be decided by a \protect\fbox{\sc {sampler}}. 
    This decision can taken in a number of ways, such as based on the collected data as depicted 
    in (a). It may also depend only on the most recently evaluated structure as depicted in (b). 
    A population-based sampling scheme is depicted in (c). Here, parent and child structures are 
    put under selection pressure where only the fittest structures are allowed to survive to the 
    next generation. The modularity of AGOX makes experimentation using different sampling strategies easy.}
    \label{fig:pictorial_sampler}
\end{figure}

The \fbox{\sc {sampler}} module provides one or more structures to be
used by other modules of the search algorithm to further exploitation
of structurally unique areas of search space, or to serve as input in the
formation of new candidates causing further exploration of non-visited
regions of search space. Depending on the implementation, the module can do so
based on all previously studied structures (or
even proposed candidates) or it can maintain a population of selected structures, Figure \ref{fig:pictorial_sampler} depicts 
three different ways that a sampler may function. In basin-hopping, the sampler would use the metropolis acceptance criterion to decide whether or not to update the sample according to the most recently evaluated candidate. 
In an evolutionary algorithm, the sampler would maintain a population of candidates that are structurally diverse 
which can be used as parents for subsequent candidates. In general, the sampler can be dependent on the order in which the data is obtained or it can be a function of the
gathered data, as is the case for the K-means sampling technique
\cite{Malthe2021} we employ for GOFEE searches in this paper. 

The \fbox{\sc {generator}} module produces new candidates either by manipulating structures taken from the sampler module or
through stochastic process that somehow proposes a set of coordinates, e.g. by placing the atoms randomly in the simulation cell.
Generators can be biased in a number of ways that may improve the performance of the search for a set of problems. 

The \fbox{\sc {collector}} module manages the generation of new candidates. In general, more than one candidate may be generated at a time
and the collector module defines how many candidates are generated using each type of generator, e.g. a preset number of each, a 
probability for each - or even as a function of the number of iterations.

The \fbox{\sc {postprocessor}} module performs actions on the generated candidates. This could be local optimization of the candidate in a machine learning model potential.
The postprocessor may also discard a candidate if it consists of
multiple unbonded fragments or if it has some unfavourable atomic arrangements, e.g.\ an
expected too short bond, an unphysical local coordination, or a too
high local energy according to a machine learning model.

The \fbox{\sc {acquisitor}} module decides which candidates are evaluated in the target potential. This concept is based on Bayesian approaches, such 
as BOSS or GOFEE \cite{Milica2019, Bisbo2020}, but it is implicitly a part of all global optimization 
algorithms as the choice of which candidates to evaluate in the target potential is central to the 
global optimization task. Some popular global optimization algorithms, 
such as BH or RSS, implicitly use an acquisitor that accepts 
all generated candidates. GOFEE on the other hand leverages Bayesian statistics 
to intelligently decide which candidate out of a collection of candidates is most likely 
to further progress the search. So, while the term \textit{acquisitor} is derived from Bayesian methods, it is in fact something all 
global optimization methods apply.

The \fbox{\sc {evaluator}} module evaluates the property of interest for the next
candidate. This may be the single candidate produced in a RSS or BH
search or the most promissing candidate in a GOFEE search.
In this work, the total energy in the target potential is the property of interest, but evaluators for other properties can be 
added such that these properties may be optimized for. Two types of evaluators are used in this work, one that 
performs a local geometry optimization used with the RSS, BH and EA algorithms and one that just performs a single-point calculation 
used with GOFEE. Another possibility is to do a limited number, $N_s$,
of relaxation steps in the target potential. In GOFEE it has been
suggested to use $N_s=1$ meaning
that a total of two single-point DFT calculations are performed, an
approach dubbed "dual-point evaluation" \cite{Bisbo2020}.

\begin{figure}
    \includegraphics[]{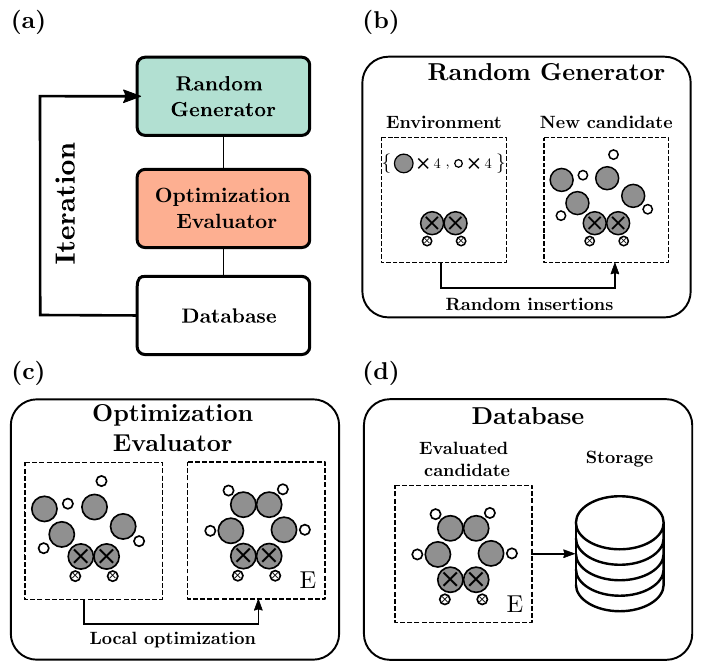}
    \caption{(a) Flow diagram of a random-structure search algorithm
      shown in terms of the involved action-type AGOX
      modules. (b)(c)(d) Illustration of the actions within the modules.}
    \label{fig:rss_agox}
\end{figure}

In Figure \ref{fig:rss_agox} we show the modules involved in a
random-structure search. Figure \ref{fig:rss_agox}(a) depicts the flow
diagram for the three action-type modules while Figure
\ref{fig:rss_agox}(b)(c)(d) details it further with the pictorial
illustrations of the modules.

The framework does not require that the modules are used in any specific order or that all of them are used. This makes the framework very flexible. 
It is enabled by an observer-pattern, which is a software design technique that will be discussed further in the next section along with the goals 
of the code. 

\subsection{Code} 
The modules described in section \ref{sec:AGOX_framework} are abstractions that we believe 
are useful when thinking about and describing atomistic global search algorithms. In order to use them in practice they 
need to be translated into code. It is, however, worth discussing briefly what makes a codebase useful. Our focus has 
been on three goals
\begin{itemize}
    \item Ease of development.
    \item Ease of use. 
    \item Performance (where necessary).
\end{itemize}
Here, ease of development covers everything from testing new ideas to fully implementing new algorithms. Ease of use 
enables new users to efficiently utilize the code even though they may not be experts in the field. Finally, any code needs to be performant, however whereas the previous points
largely go hand in hand, performance can come at the cost of increased complexity. For many atomistic search problems, 
the most time-consuming step will be the quantum mechanical calculation of the total energy in the target potential, e.g. density 
functional theory (DFT). In order to leverage the many efficient electronic structure codes available, AGOX uses the atomistic simulation
environment package \cite{Larsen2017} (ASE) which has Python interfaces to a large collection of codes. Compared to the time of a 
DFT calculation, the generation of a candidate structure is very fast, even without extremely optimized code. Therefore, for 
such modules it is more important that the code is easy to understand
and fast to develop further. Even though local optimization in a machine learning 
potential is many orders of magnitude faster than in DFT, the
computational time spent in optimizing many candidates may become
comparable to that spent in
a single-point DFT calculation. Therefore, it is worth some additional complexity to ensure that such optimizations are 
done efficiently. 

\begin{figure}
    \centering
    \includegraphics[width=0.5 \textwidth]{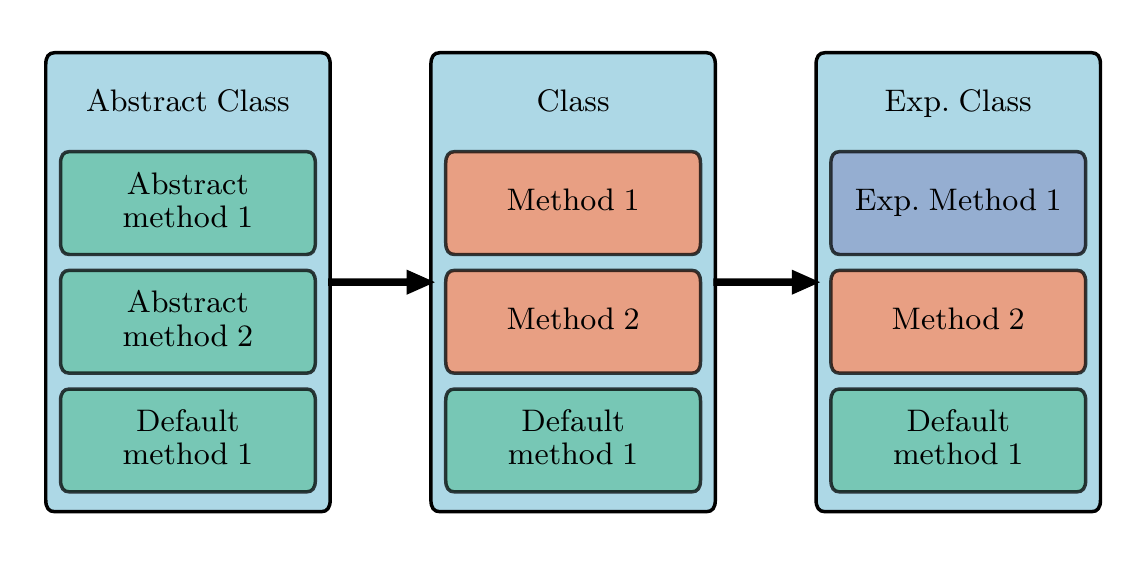}
    \caption{Illustration of inheritance from an abstract base class shown on the left. The 
    ABC defines two abstract methods and a single default method, a specific version of the class (middle) 
    requires real implementations of the two abstract methods. Experiments with the functionality of this
    specific version can be performed by inheriting from the it and surgically replacing only the relevant 
    method.}
    \label{fig:OOP_inheritance}
\end{figure}

In order to ensure both ease of use and ease of development, we take advantage of the objective-oriented programming (OOP)
capabilities of the Python language. An abstract base class (ABC) has been implemented for each of the modules of the framework. 
These are essentially code templates that define the required methods and attributes of an actual specific implementation of 
any of the modules. The ABC may also define a number of functions that are generally convenient for that particular module, 
such as checking whether a set of coordinates have chemically valid bond lengths. This improves both the readability of the 
code by hiding the details behind a method-call when those details are not necessary and the reliability by only having one common implementation for each such method. 
The OOP design also allows us to leverage inheritance when developing new features in two ways. The first is as mentioned 
when a new version of one of the modules of the framework inherits from the ABC and the second when an implementation 
inherits from another specific implementation. Inheritance is depicted in Figure \ref{fig:OOP_inheritance}.

To make the framework as flexible as possible, we do not wish to impose any restrictions on how the modules are used to design a search algorithm.
This means, that it has to be possible to leave out a module, change the order of modules and use multiple versions of the same module. 
This is accomplished by building an observer-pattern into the core iterative loop \cite{Gamma1994}. An observer-pattern is a software design technique that entails a 
subject maintaining a list of its observers that it will notify when changing its state. In AGOX, this state is the iteration counter and 
observers, which are AGOX action-type modules, will be notified in a pre-specified order in each iteration. By changing the list of observers one may alter what the program actually does. Importantly, this 
can be done without altering the code of any of the modules or the core iterative loop. Each module acts as an observer that can read and write to a shared data collection, 
such that there is no hardcoded order of execution. A valid AGOX program can range from a for-loop that does nothing each iteration to a GOFEE search and beyond. 

\begin{figure}
    \includegraphics[width=0.45\textwidth]{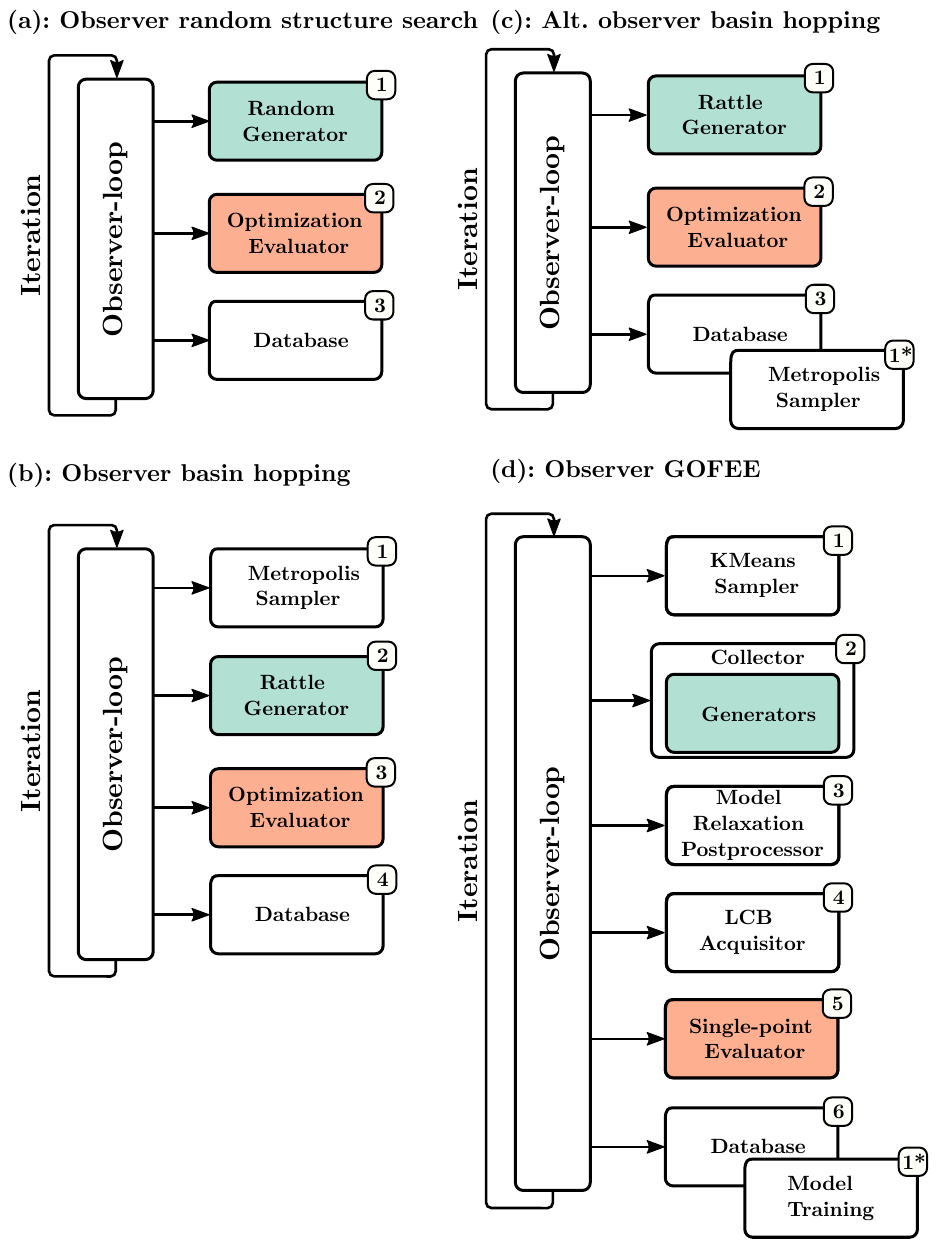}
    \caption{Several global optimization algorithms programmed in different ways. The observer-pattern loop allows different 
    algorithm to be expressed using the same modules. The numbers note the order of execution, which the observer-pattern takes into 
    account. A module may also be an observer to the database and be notified whenever new data is added. This is depicted for the sampler module of 
    basin-hopping and the model training function in GOFEE.}
    \label{fig:observer_patterns}
\end{figure}

\begin{figure}
    \includegraphics[width=0.40\textwidth]{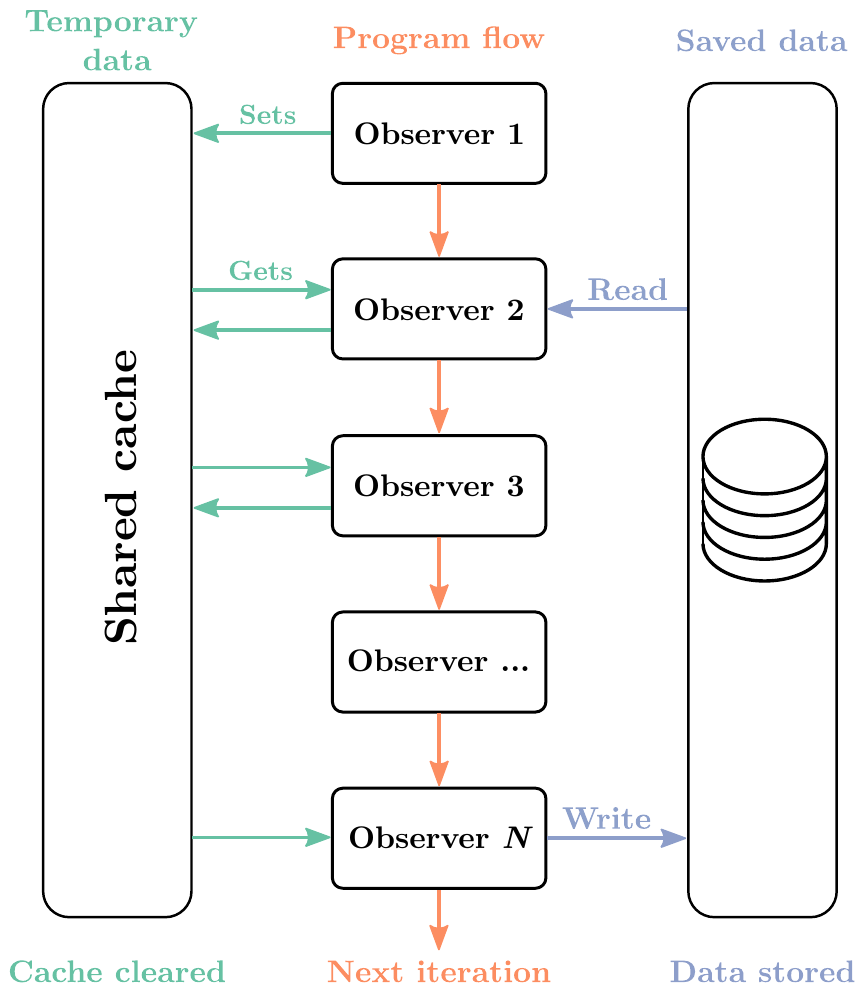}
    \caption{Depiction of the program flow and data access of an AGOX algorithm. Here, an abstract 
    algorithm is depicted as a series of observers, each of which have the ability to get or set data 
    in the shared cache. Additionally, modules can have access to the database. At the end of an 
    iteration the shared cache is cleared, whereas data saved to the database is saved permanently. 
    Observers may also communicate with each other through connections
    defined in the AGOX python script, whenever modules depend on
    other modules to perform their function. For instance, a generator may be given 
    access to the sampler in order to retrieve sample members.}
    \label{fig:fig_cache}
\end{figure}

Examples of such observer-pattern based algorithms are depicted in
Figure \ref{fig:observer_patterns}. The RSS method, shown in Figure
\ref{fig:observer_patterns}(a), has three observer-type modules attached
to the iteration loop. Similarly, the BH method may be implemented
with four observer type modules	attached to the iteration loop, as
shown in Figure \ref{fig:observer_patterns}(b). Since the observer-pattern is
also built into the \fbox{\sc {database}} module, an alternative
implementation of the BH methods can be layed out as shown in Figure
\ref{fig:observer_patterns}(c). Here, the  \fbox{\sc
  {sampler}} module is moved from being an observer on the main
iteration loop to being an observer on the \fbox{\sc {database}}
module. In this implementation, the \fbox{\sc                                                                  
  {sampler}} module is invoked whenever new DFT-level data has been
dealt with by the \fbox{\sc {database}} module. The observer pattern
on the \fbox{\sc {database}} module is also exploited in the
implementation of the GOFEE method presented in Figure
\ref{fig:observer_patterns}(d). Here, it is the update of the
\fbox{\sc {model}} that is invoked whenever the \fbox{\sc {database}}
module has handled new DFT level data.
To allow decoupling of modules, such that e.g.\ a \fbox{\sc {postprocessor}}  does not expect input from a \fbox{\sc {generator}} 
or any other instance of hard coded interdependence, the communication between modules is handled by a shared cache 
or through the database, as depicted in Figure
\ref{fig:fig_cache}. 

A useful way of thinking about an AGOX program is to consider it a set of functions  
that are executed in an order specified in a script, rather than by the underlying core code and without any predefined relations between the set of functions.  
AGOX is designed to work based on definition and assembly of such
functions via definition of modules at the scripting level since it
provides an easy and logical way of handling algorithms deficient of
certain actions. Imagine, by the contrary, that connections between the
modules were indeed hardcoded to e.g.\ match the GOFEE layout
presented in Figure \ref{fig:agox_flowchart}, then the RSS algorithm 
could still be formulated having versions of certain modules that
would do essentially nothing, e.g. the acquisitor, the sampler and the
postprocessors. With the observer-pattern 
these modules can be left out entirely. Another advantage granted by
the observer-pattern is that new modules may be included in the future, again because the 
order of execution and the communication between modules is not predefined.

\subsection{Code example: An acquisitor}
As an example of the benefits of the described coding scheme, we will take a look at an acquisitor. The basic acquisitor used 
for GOFEE type searches is the lower confidence bound (LCB) acquisitor, see Section \ref{sec:GPR} for details.
The implementation of LCB inherits from the acquisitor ABC, in the way depicted in Figure \ref{fig:OOP_inheritance}.
We can easily experiment with alternative acquisition functions, and depending on the type, it can either inherit from the 
base-class or a specific implementation, e.g. the LCB acquisitor. As an example, we can imagine we would like to try an 
acquisition function defined by 
\begin{equation}
    x_a = \argmin_{x \in X} \left[ E(x) - \kappa \sigma(x)^\rho \right],
\end{equation}
where $X$ is a set of candidate coordinates, $E(x)$ and $\sigma(x)$ are surrogate energy and uncertainty functions and $\kappa$ and $\rho$ are chosen parameters. 
For $\rho=1$ this reduces to the LCB expression. This can be implemented in
very little code by inheriting from the LCB class. 
\begin{lstlisting}[language=Python]
class PowerLowerConfidenceBoundAcquisitor(LowerConfidenceBoundAcquisitor):

name = 'PowerLowerConfindenceBoundAcquisitor'

def __init__(self, rho=1, *args, **kwargs):
    super().__init__(*args, **kwargs)
    self.rho = rho

def acquisition_function(self, E, sigma):
    return E - self.kappa * sigma ** self.rho

def acquisition_force(self, E, F, sigma, sigma_force):
    return F - self.kappa * self.rho * sigma**(self.rho-1) * sigma_force
\end{lstlisting}
Thus, in less than 15 lines of code, experiments can be made with a different acquisition function without any risk 
of causing issues with existing code due to the use of inheritance. 

\subsection{Success curves}

\begin{figure}[h]
    \centering
    \includegraphics[width=0.45\textwidth]{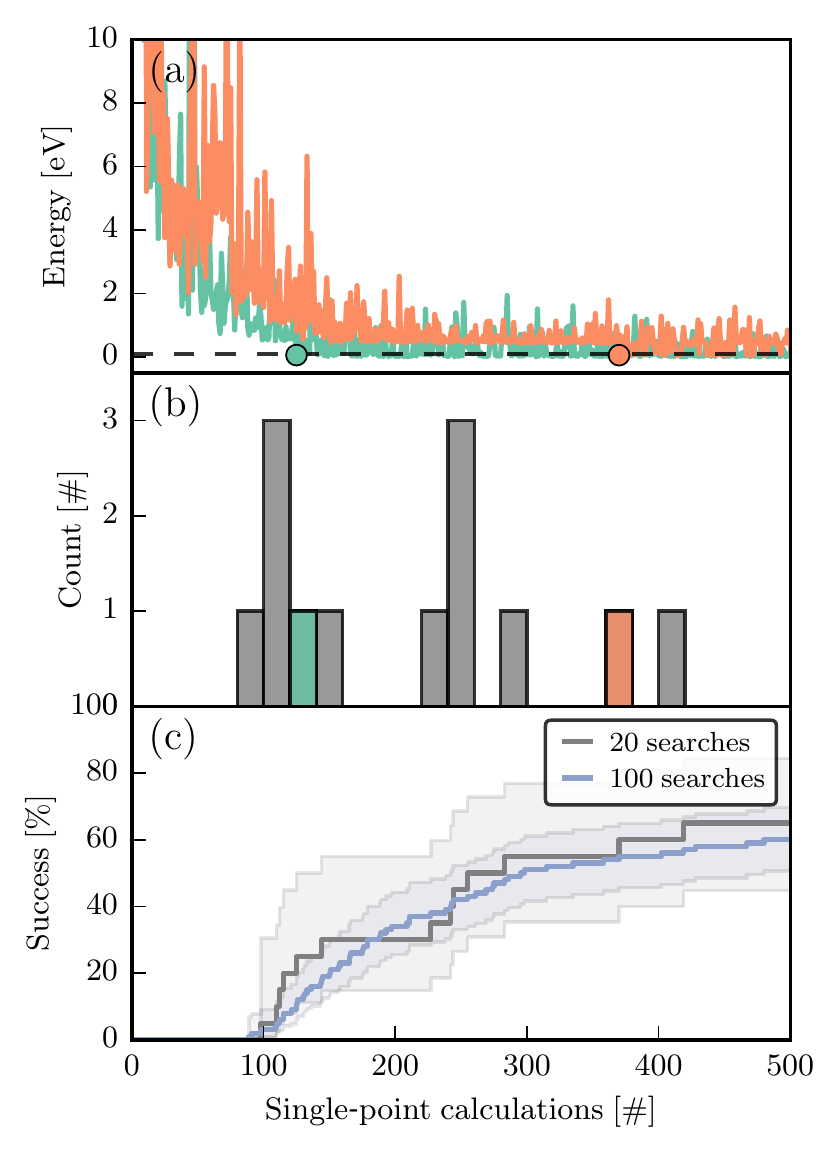}
    \caption{(a) The energy per single-point calculation for two different searches, the 
    dashed black-line indicates the success criterion. The two-colored dots show when 
    the two searches find a candidate that is considered successful for the first time. 
    In (b) the number of single-point calculations until first success is presented 
    as a histogram, green and orange bars indicate the contributions from the searches in (a). 
    A success curve is obtained by integrating the histogram (with a bin size of 1) and normalizing 
    according to the number of searches, as shown in (c). With more searches the curve becomes smoother and 
    the uncertainty decreases as evidenced in the difference between the gray curve based on 20 searches and the blue based on
    100 searches in (c).}
    \label{fig:succ_how_to}
\end{figure}

One key metric to judge the performance of any global optimization algorithm is a success curve, which is a statistical 
property that measures the percentage of independent search runs or searches that are 'successful' against the number of single-point calculations, iterations or timing metrics. 
In the used terminology, a search refers to the execution of an algorithm typically for a predetermined number of iterations - although other stopping criteria 
are also possible. A search produces a number of structures, e.g. a
random-structure search of 1000 iterations will produce 1000, not
necessarily unique, local minimum energy structures. 
However, given the stochastic nature of most global optimization algorithms there is no guarantee that a search finds the global minimum structure. 
Whether or not a single search finds the global minimum also tells us very little about the ability of the algorithm to solve the problem, 
again due to the stochastic nature, a second search may 
lead to different results. To compare searches done with different algorithms, or the same algorithm with different parameters, we therefore 
need a statistical measure that averages over several independent searches, which is exactly what a success curve does. 
Since the underlying physics is typically not only governed by the
global minimum energy structure, but by a collection of low-energy structures, we count
structures within some small energy  of the best structure found among all searches to be successful in this work. For applications where finding a specific structure is considered
the success criterion, a graph-based method for identifying that
specific structure can be employed, see Sec.\ \ref{sec:graph_spectral}. With either success criterion, success curves
are informative about the performance of the used algorithm. As such, they can be used as a measure of confidence in the found solution,
with high success being required in order to be confident in the solution and the ability of the method to solve more difficult problems. 
The process of obtaining a success curve is illustrated in Figure \ref{fig:succ_how_to}. 

\section{Application: Global optimization algorithms}
\label{section:pt14}
As an example of an application of the AGOX framework, we study a system consisting of a platinum cluster on a gold surface 
described by the simple effective medium theory (EMT) potential \cite{Jacobsen1987} as implemented in ASE. This potential is chosen as it 
works out of the box having installed just ASE and AGOX. We present the results of the application of the AGOX framework using four 
different search algorithms, namely RSS, BH, EA and GOFEE. 

In order for an optimization algorithm to be efficient at solving any particular problem it must impose search biases 
that favour finding the global optimum solution for that particular problem. We distinguish between two kinds of 
biases. Those that are imposed from the outset and those that are learned by the algorithm. A somewhat trivial example of 
an imposed bias is the number of atoms, having decided that only 14 Pt atoms are present, the search space is limited to solutions 
that involve 14 Pt atoms. A less trivial imposed bias, is to constrain the physical space that the algorithm is allowed to use, 
this is useful as it limits the number of symmetry related solutions, e.g. translations of the cluster along the
surface. Learned biases arise from how the algorithm uses the data it gathers during a search.
The surrogate potential will impose a learned bias and so will the sampling technique used to decide which previous 
structures are used to generate new candidates. 

\begin{figure}
    \centering
    \includegraphics{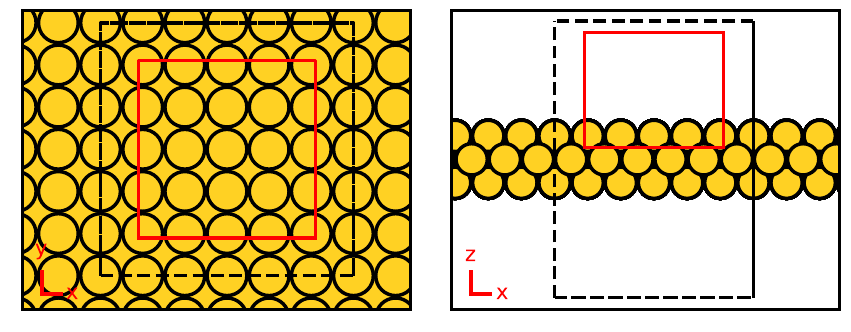}
    \caption{Gold surface slab used as template for the Pt$_{14}$/Au(100)
      search. The computational super cell and confinement cell are
      depicted in the $xy$- and $xz$-planes in black and red, respectively.}
    \label{fig:Pt14Au_confinement}
\end{figure}

In AGOX, the search can be confined to a cell which may differ from the periodic/computational cell used, such that both the generators and the relaxation postprocessing will 
not result in any geometries that have atoms outside of the specified cell. For the Pt$_{14}$/Au(100) system this is essential to 
find physical solutions where Pt atoms are only present on one side of the surface slab. This cell is depicted along with
the surface slab in Figure \ref{fig:Pt14Au_confinement}. 

The system-dependent choices, that is the decisions that define the search problem, can be summarized as
\begin{itemize}
    \item The number of and species of the atoms that are directly involved in the search. (The 14 Pt atoms)
    \item The template, that is the number and positions of atoms already present in the cell. (The gold surface and cell depicted in Figure \ref{fig:Pt14Au_confinement})
    \item The confinements, if any. (The confinement cell depicted in red in Figure \ref{fig:Pt14Au_confinement})
\end{itemize}
The specific choices made here are listed in parenthesis.
Each global optimization algorithm has an additional number of
options. The action-type observer modules used in the four methods
were set up in the following way and with the following choices for
the adjustable parameters:

\subsection*{Random-structure search}

\fbox{\sc {Generator}}: One candidate is generated per iteration using the random generator. \\

\fbox{\sc {Evaluator}}: The generated candidate is relaxed in the EMT potential until the forces on all Pt atoms are below 0.05 eV/Å, 
template atoms are fixed, and Pt atoms are not allowed to leave the confinement cell depicted in Figure \ref{fig:Pt14Au_confinement}. 

\subsection*{Basin-hopping}

\fbox{\sc {Sampler}}: A new structure is accepted or rejected using
the Metropolis criterion with probability of acceptance given by
\begin{equation}
    A = \min \Big\{1, \exp{[\beta (E_{k-1}-E_k)]}\Big\},
    \label{eq:metropolis}
\end{equation}
With $\beta = 1 / k_B T$ with $k_B T = 1 \ \mathrm{eV}$ and $E_k$ is the energy of the structure found in iteration $k$. 
If accepted, the structure replaces the previously accepted structure as the starting point of 
the rattle generator in the next iteration. 
Skipped in the first iteration. \\

\fbox{\sc {Generator}}: Initialized with a random generator, and all subsequent iterations generate one candidate per iteration using a rattle 
generator applied to the latest structure accepted by the sampler. \\

\fbox{\sc {Evaluator}}: Same as for RSS. \\

\subsection*{Evolutionary Algorithm}

\fbox{\sc {Sampler}}: A population of 10 structures with diversity enforced using fingerprint feature and parent selection using 
the algorithm proposed by Vilhelmsen and Hammer \cite{Vilhelmsen2014}. \\

\fbox{\sc {Generators}}: Initialized with 10 random generator candidates and with subsequent iterations using 10 candidates generated 
by rattling structures from the population.  \\

\fbox{\sc {Evaluator}}: Same as RSS for all 10 candidates generated per iteration. \\

\subsection*{GOFEE}

\fbox{\sc {Sampler}}: We use the K-means clustering based sampling method reported by Merte et al. \cite{Malthe2021} with a total 
sample size of 10 structures and an energy requirement such that considered structures are within 25 eV of the best 
structure discovered so far. The method works similarly to the depictions in Figure \ref{fig:pictorial_sampler}(a) by applying 
K-means clustering to all evaluated candidates in feature space and selecting structures from each cluster enforcing diversity 
of the selected structures. \\

\fbox{\sc {Generators}}: We use a random generator, a rattle generator and a generation mechanism 
described by Palecio et. al. \cite{Behler2020} that favours perturbing atoms far from the center of geometry of the cluster,
which we call center-of-geometry generator. A total of 30 candidates are generated per iteration
with 10 candidates from the random generator, 15 from the rattle generator, and five from the center-of-geometry generator. \\

\fbox{\sc {Postprocessing}}: All candidates are relaxed in the LCB expression evaluated using the surrogate GPR model until the 
maximum LCB force ($\kappa=2$) of non-template atoms is below 0.2 eV/Å. This is done in parallel utilizing all available CPU cores.
The relaxation is constrained such that atoms are kept within the confinement cell. \\

\fbox{\sc {Acquisitor}}: Among the 30 candidates the one with the lowest lower
confidence bound  ($\kappa=2$) value is chosen. \\

\fbox{\sc {Evaluator}}: One single-point calculation is done for the
candidate picked by the acquisitor. \\

\fbox{\sc {Model}}: Details of the GPR model are given in Sections \ref{sec:GPR} and \ref{sec:fingerprint}.\\

\subsection*{Results}

\begin{figure}[h]
    \centering
    \includegraphics{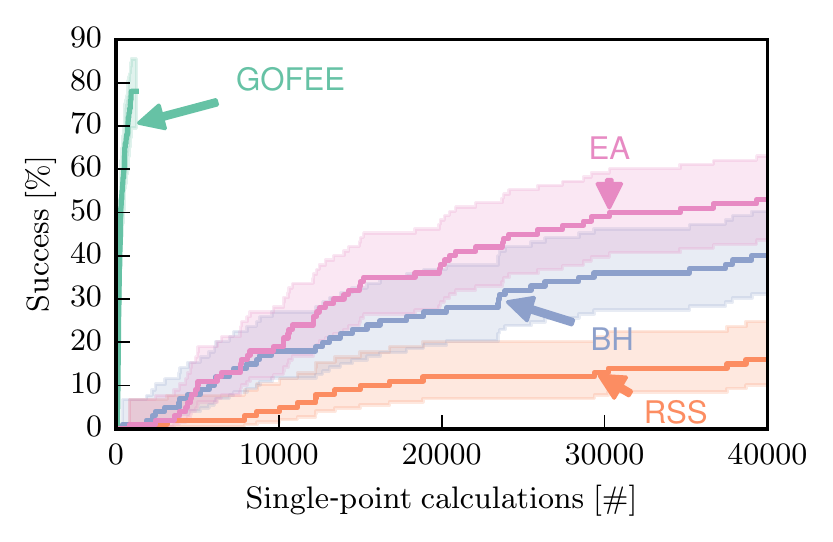}
    \caption{Success curves for four different global optimization algorithms. }
    \label{fig:Pt14_succss_example}
\end{figure}

Four different success curves are presented in Figure \ref{fig:Pt14_succss_example} that originate from sets of AGOX searches using 
the four different search algorithms, RSS, BH, EA and GOFEE. The least biased algorithm, random structure 
search, explores the search space in the least directed way which means it will not become stuck but at an increased 
computational cost. Basin-hopping, that reuses previous structures in the generation mechanism, is directed towards 
exploring low-energy regions which results in higher success with fewer single-point calculations. The EA is able to 
evolve a number of structures, making it more resistant towards getting stuck in a local minimum and enabling slightly 
better performance compared to the BH in agreement with similar studies \cite{Bauer2022}. Note that the EA could also use a crossover generation
mechanisms that would involve combining two or more members of its population, but that is not taken advantage of here. 
Finally, the GOFEE algorithm replaces local optimization in the true potential (EMT) with local optimization in a
surrogate GPR model and only does single-point calculations for favorable structures chosen by the LCB acquisition 
function. This results in orders of magnitude fewer single-point calculations while reaching much higher success. 
It should be noted that the parameters of each algorithm have not been optimized, as that requires a very large 
computational effort as a single success curve requires at least tens of searches, likely more to resolve subtle differences for 
fairly unresponsive parameters. Regardless, the objective here 
is to showcase that all of the algorithms can be handled within the AGOX framework. Figure \ref{fig:Pt14_succss_example}
does show that the GOFEE algorithm heavily outperforms the other three algorithms in terms of single-point calculations, 
without the prospect of the others improving enough with optimal
parameters to compete with it. 

This is not surprising given that GOFEE only performs one single-point calculation per iteration while relying on the 
surrogate model to get locally optimized structures, which is an algorithmic change rather than due to 
specific parameter choices. As an EMT potential is used here, it is in fact more computationally demanding to run 
GOFEE for this problem, as the surrogate potential is not faster to evaluate than the potential, but GOFEE needs to 
query the potential much fewer times which is the key benefit for problems with a more computationally demanding 
potential. \\

\begin{figure}[h]
    \centering
    \includegraphics[width=0.45\textwidth]{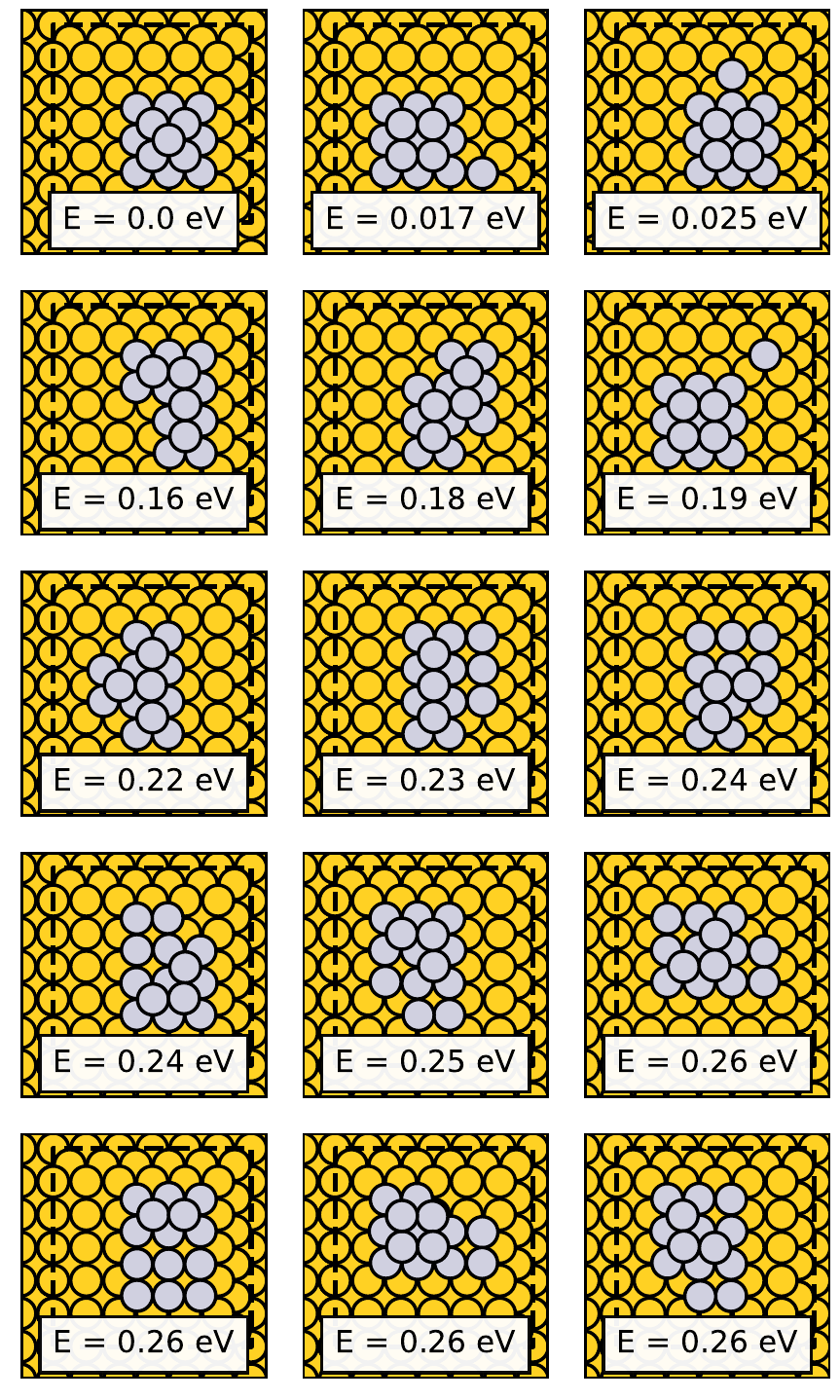}
    \caption{Low-energy structures of Pt$_{14}$ on Au(100), with the energies relative to 
    the structure with the lowest energy. The structures have been locally optimized after the search such that the maximum force is 0.01 eV/Å.}
    \label{fig:Pt14_best_structures}
\end{figure}

In Figure \ref{fig:Pt14_best_structures} we report the 15 structures
with the lowest energy for the Pt$_{14}$/Au(100) system.
These have been extracted by first obtaining all structures found that are within 0.5 eV of the best structure.
Among these, those that have unique graph spectra, according to the method described in Sec. \ref{sec:graph_spectral}
are identified and locally optimized, after which the graph spectra are compared once again. It is apparent, that the methods
are capable of finding many distinct low-energy structures and that the
graph-based sorting criterion enables the distinction of
the different low-energy structures.

\section{Application: Machine learning-enhanced basin-hopping}

As an example of the ability of AGOX to facilitate the design of global optimization algorithms, we present a comparison 
of a basin-hopping algorithm and a machine learning-enhanced
basin-hopping algorithm. As discussed previously, the most 
expensive part of a basin-hopping type search is the local optimization, often involving many dozens of electronic structure 
calculations to obtain the forces on each atom. If some of these calculations can be omitted while retaining
the ability of the algorithm to solve any given problem that would constitute a more computational efficient algorithm. \\

\begin{figure}
    \centering
    \includegraphics[]{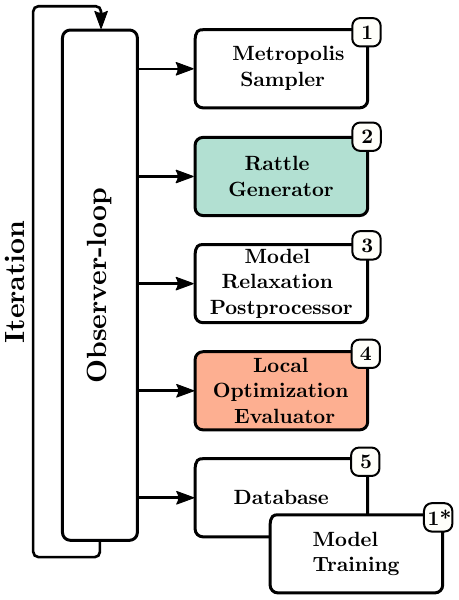}
    \caption{Basin-hopping search algorithm with a model inserted.}
    \label{fig:model_basin_hopping}
\end{figure}

In AGOX such algorithmic changes can be made easily and surgically
requiring only changes to the script and not to the core of the code. These specific changes involve inserting a surrogate model local optimization 
step between the steps performing candidate generation and candidate
optimization as for BH in Figure \ref{fig:agox_flowchart} or in the terminology of AGOX 
a postprocessor in between the rattle-generator and local optimization
evaluator observers of Figure \ref{fig:observer_patterns}(a). The resulting
flowchart of the algorithm is shown in Figure \ref{fig:model_basin_hopping}.
The changes compared to a standard basin-hopping search script amount to, defining a model 
\begin{lstlisting}[language=Python]
    model = ModelGPR.default(environment, database)
\end{lstlisting}
and giving that model to a postprocessor
\begin{lstlisting}[language=Python]
    relaxer = RelaxPostprocess(
    model=model, start_relax=10, 
    optimizer=BFGS, optimizer_run_kwargs=
        {'fmax':0.05, 'steps':100},
    constraints=environment.get_constraints())
\end{lstlisting}
and giving those additional modules to the AGOX class
\begin{lstlisting}[language=Python]
    agox = AGOX(database, generator, sampler, evaluator, relaxer, wrapper)
\end{lstlisting}
The scripts are available in full at \agoxdata. \\ 

\begin{figure}
    \includegraphics{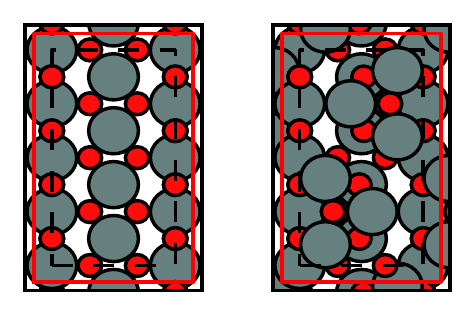}
    \caption[]{Left: Template comprising a one-layer rutile
      SnO$_2$(110)-(4$\times$1) slab. Right: Global minimum energy structure
      for Sn$_6$O$_6$ on SnO$_2$(110)-(4$\times$1). Sn atoms
      are shown in gray, oxygen atoms are shown in red. The super cell
      is drawn as a dashed black rectangle. The confinement in the
      $xy$-plane is indicated as a red colored rectangle.}
    \label{fig:sno2_config}
\end{figure}

We applied the ML-assisted basin-hopping algorithm to optimizing a
rutile SnO$_2$(110)-(4$\times$1) system where 6 Sn and 6 O atoms are arranged on the fixed surface described at the DFT level using an LCAO basis set \cite{GPAW_LCAO}
and the PBE exchange-correlation functional
\cite{perdew1996generalized}
as implemented in GPAW
\cite{Mortensen2005, Enkovaara2010}. The $\Gamma$-point is used for
sampling the Brillouin zone and for Sn a 4-valence electron PAW setup is used.
The system is shown in Figure \ref{fig:sno2_config}. The settings of
the modules were as follows:

\subsection*{ML-enhanced basin-hopping}

\fbox{\sc {Sampler}}: Same as for BH in Section \ref{section:pt14}. \\

\fbox{\sc {Generator}}: Same as for BH in Section \ref{section:pt14}.\\

\fbox{\sc {Postprocessor}}: Invoking the BFGS optimizer in ASE using
the model calculator for up to 100 episodes or until the forces of
non-template atoms are below 0.05 eV/Å. \\

\fbox{\sc {Evaluator}}: Same as for RSS in Section \ref{section:pt14}, except that only a limited
number of relaxation steps were done ($N_s=3$ or $N_s=10$). \\

\fbox{\sc {Model}}: Same as for GOFEE in Section \ref{section:pt14}. \\

\begin{figure}
    \includegraphics{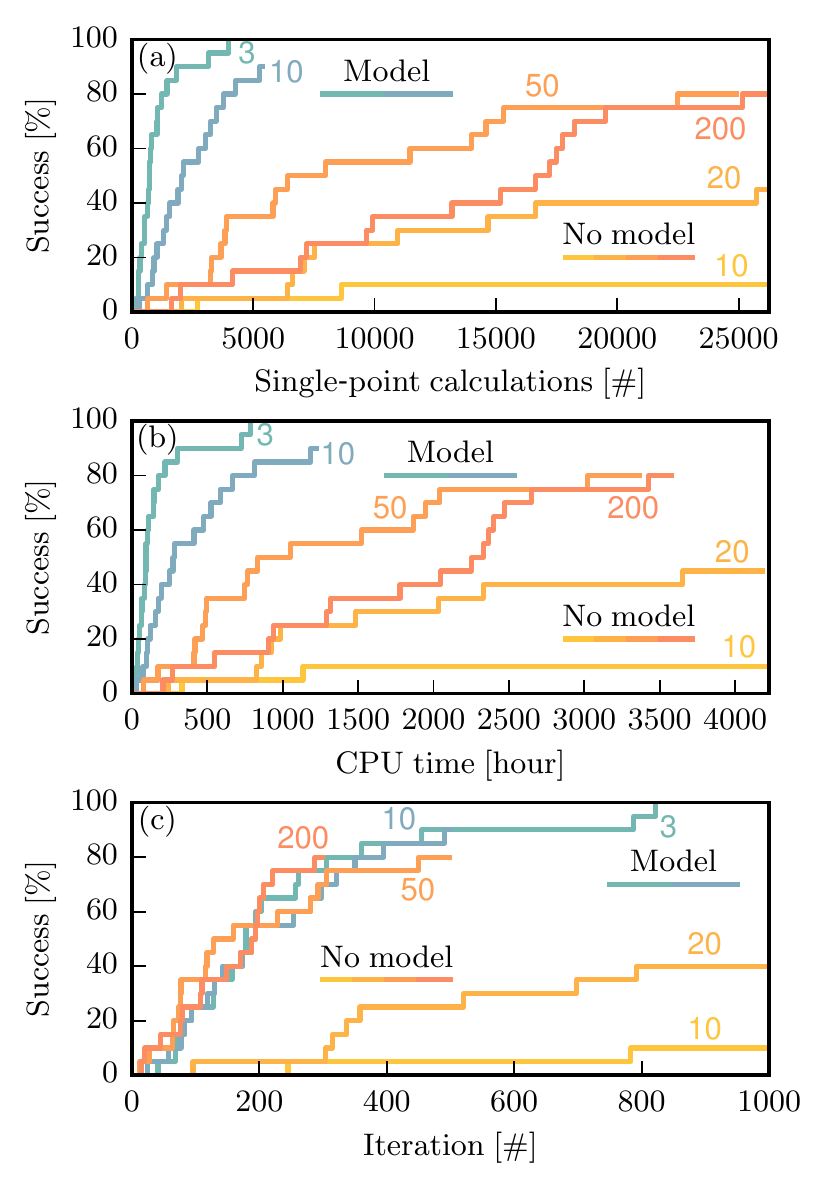}
    \caption{Success measured against (a) single-point DFT
      calculations, (b) CPU time and (c) number of basin-hopping iterations. 
    The colored numbers indicate the maximum number of DFT gradient
    steps allowed in each basin-hopping iteration. The
    CPU time is derived as 24 times the wall time passed, since the runs were
  performed on 24 CPU cores in order to speed up the DFT calculations
  that parallelize well. In (c) the iteration axis has been limited to 1000, the standard basin-hopping searches performing 10 
  and 20 relaxations steps have run for a total of 2700 and 1400 iterations, respectively.}
    \label{fig:sno2_success}
\end{figure}

In Figure \ref{fig:sno2_success}, we show success curves for solving this problem with and without locally optimizing in the 
model prior to doing DFT optimization. For both methods searches, are done with a different maximum allowed number of 
DFT gradient steps, as that is an important parameter for the ability of standard basin-hopping to solve a problem.
In Figure \ref{fig:sno2_success}(a) we see that using a model much fewer DFT steps are required, which is the same trend observed for the previous system. As the DFT calculation 
time is the dominating factor a similar trend is observed when plotted against CPU time in Figure \ref{fig:sno2_success}(b), however the curves 
using a model are slightly less steep because of the scaling of training and using the model. Finally, in Figure \ref{fig:sno2_success}(c)
the success is plotted against the number of basin-hopping iterations, presenting a remarkable result that 
even though the model is learning the potential on-the-fly, there is no difference in the number of iterations 
required to solve the problem. Developments such as the one presented here are enabled by AGOX as algorithms can 
be altered entirely through a script.

\section{Application: Parallel tempering}
An important consideration when building software tools is the ability to take advantage of the ever-increasing number of processor cores in 
modern computers, be it a desktop computer with a handful of cores or a high-performance computing server with many thousands. 

Within a single search run AGOX takes advantage of ASE to run electronic-structure calculations in parallel on the number of processes allotted. Furthermore,
in the GOFEE setting, where several candidates are produced per iteration, model relaxations are also parallelized such that the same number 
of cores are utilized as for electronic structure calculations. 

In general, it is also an advantage to perform several instances of the same search with the same settings in order to have several independent search runs
of the algorithm, in order to more thoroughly explore the search-space. Due to the independence of each search, with no communication to any other searches, this is an \textit{embarrassingly} parallel task, 
and parallelization is as simple as running the same script on several
computers - or using a workload manager, such as Slurm commonly available on HPC facilities. 
This is what is done to make success-curves that require a large number of independent searches as discussed previously, and what we recommend doing when applying AGOX to solve a search problem.

AGOX can, however, also be used to build algorithms that benefit from
having several workers. An example of such an algorithm is parallel
tempering \cite{Kofke2002}. In parallel tempering, 
basin-hopping searches are performed simultaneously at different temperatures. Structures accepted at different temperatures
may be swapped to avoid stagnation by promoting exploration at high temperatures and exploitation at low temperatures. In this setting, 
a search consists of a number of workers using different processors with each worker running a basin-hopping search at one temperature and all workers sharing a 
single \fbox{\sc {database}}.

Structures are swapped between workers with adjacent temperatures every $N_t$ episodes with probability 
\begin{equation}
    P = \min \Big\{1, \exp{(\beta_i}-\beta_j)(E_i-E_j)\Big\}
    \label{eq:parallel_tempering}
\end{equation}
where $\beta_i = 1/k_BT_i$. Temperatures are chosen according to $k_BT_i = k_BT_0 \cdot (3/2)^i$ where $i$ ranges from 
zero to the total number of workers $N_w$ minus one in integer steps and $k_BT_0 = 0.05 \ \mathrm{ eV}$. 
Every $N_t$ iterations each worker waits for all other workers of the same search to reach that iteration before reading the database from disk, 
this is done prior to swapping structures to ensure all workers are synchronized. Compared to a standard basin-hopping search this type 
of parallel tempering search requires only changing the database module and the sampling module.

\subsection*{Synchronized parallel tempering}

\fbox{\sc {Sampler}}: Parallel tempering using the Metropolis criterion of Eq. \eqref{eq:metropolis} to decide 
whether or not to accept a candidate and attempting swaps between candidates at adjacent temperatures every $N_t=10$ episode according to Eq. \eqref{eq:parallel_tempering}. \\

\fbox{\sc {Generator}}: Same as for BH in Section \ref{section:pt14}. \\

\fbox{\sc {Postprocessor}}: Generated structure is moved to the center of the cell. \\

\fbox{\sc {Evaluator}}: Local optimization until forces on all atoms are below 0.2 eV/Å. \\

\fbox{\sc {Database}}: Structures are synchronized among workers every $N_t$ iteration. \\

\begin{figure}
    \includegraphics[]{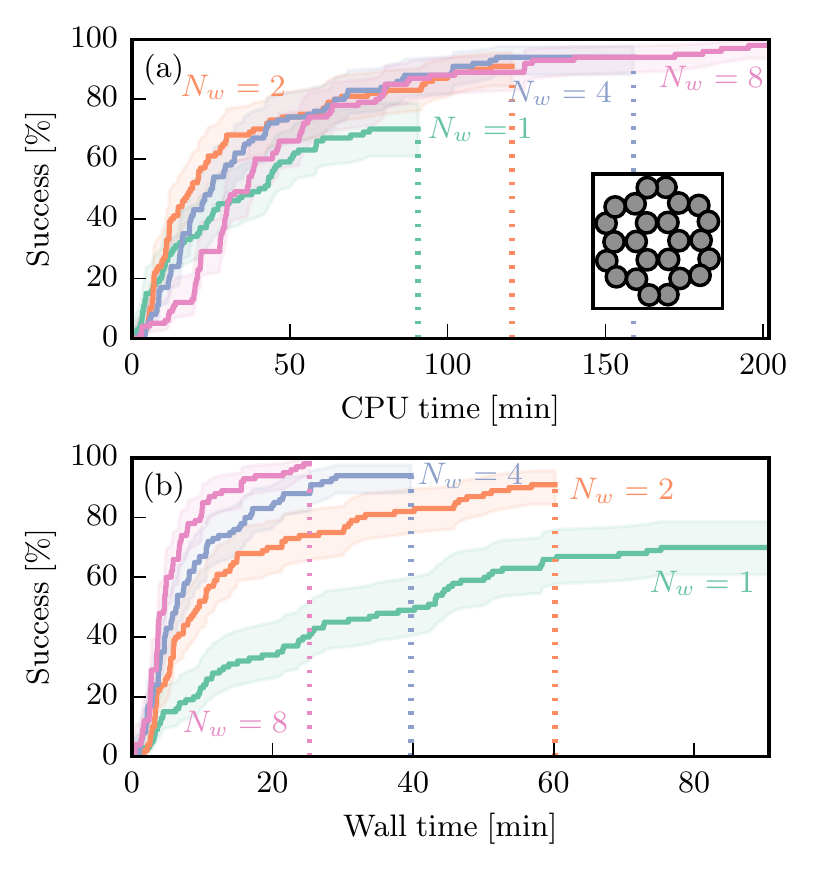}
    \caption{Success curves for parallel tempering with $N_w = \{1, 2, 4, 8\}$ as a function of 
    (a) CPU time and (b) wall time. In both cases it is the time per independent search. The inset in 
    (a) shows the global minimum structure.}
    \label{fig:parallel_tempering}
\end{figure}

As an example, we apply this algorithm to a 24-atom carbon cluster constrained to two-dimensions in the search described by the semi-empirical extended tight-binding method (GFN2-xTB) \cite{xtb_gfn2, xtb_review}. 
To fairly compare searches with different number of workers each search is run for $2000 / N_w$ iterations, such that the same number of candidates are generated and locally optimized. 
Success curves for this system are shown in Figure \ref{fig:parallel_tempering} both as a function of CPU time and wall 
time in (a) and (b) respectively. The parallel-tempering scheme helps alleviate stagnation, as evidenced by searches with more workers reaching 
higher success rates for the same amount of CPU time. The overhead introduced by workers having to wait for each other to synchronize the 
database does make it more expensive in CPU time to do searches with more workers. However, as each worker only has to do a fraction of the work, the wall
time, that is the waiting time between starting search runs and achieving results, is decreased significantly. 

This example show-cases one example of how AGOX modules may be used to parallelize algorithms. The shared database 
module can also be used to parallelize other algorithms implemented in AGOX due to the modularity of the framework. 
That could for example be random structure search runs, which parallelize trivially, or the model-enhanced basin-hopping 
example of the previous section, where the model for each search may be updated based on the collected data from all searches.

\section{Application: Embedded metal cluster}

\begin{figure}[h]
    \includegraphics{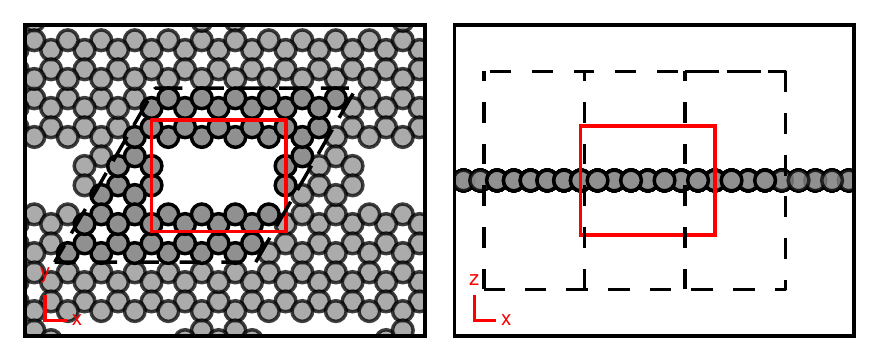}
    \caption{Template used for Ru$_3$N$_4$C$_4$ with the confinement cell shown in red.}
    \label{fig:RuN_template}
\end{figure}

Finally, we have applied AGOX to a system that consists of three Ru, four Ni and four carbon atoms that are embedded in a hole in a graphene 
sheet. This system is inspired by a combined experimental and
theoretical study of the properties of such
graphene embedded Ru$_3$N$_4$ clusters \cite{Shufang2017}. Here, we only employ the GOFEE algorithm as described in Section \ref{section:pt14}, except the 30 candidates per iteration are generated only with a random generator and a 
rattle generator, producing 10 and 20 candidates, respectively. The choices of parameters and analysis procedure are not 
specific to GOFEE and can be used with any global optimization algorithm. Only the three Ru, four Ni and four carbon atoms are allowed to move 
during the search. To accurately rank the most stable structures, we apply a procedure where all atoms are 
involved in a local optimization after the global optimization algorithm has finished, this procedure is described in detail 
below. 

The template and confinement used for the search 
is illustrated in Figure \ref{fig:RuN_template}, the periodic cell is 16 \AA \ in the $z$-direction. For this system the search is performed 
at the DFT level using GPAW \cite{Mortensen2005, Enkovaara2010} with a plane wave basis set using an energy cutoff of 300 eV and only a 
single k-point with the Perdew-Burke-Ernzerhof (PBE) exchange-correlation functional \cite{perdew1996generalized}. Further analysis 
of the found structures is performed with a plane wave energy cutoff of 400 eV and a (3, 3, 1) Monkhorst-Pack k-point grid
again with the PBE functional, but now also including the effects of spin polarization. Henceforth, we call the first set of 
settings the 'rough' settings and the latter set the 'fine' settings. 

For studying the physical properties of a system using DFT, it is important that the computational settings are chosen at
a sufficient level of precision in order for the relevant properties to converge. This is typically done by performing 
convergence checks were each setting is varied until the property of interest is converged within a specified tolerance. 
For global optimization, the target property is the geometry of low-energy structures, but running the entire search with 
multiple sets of settings is computationally costly so that cannot generally be done. Furthermore, it is often the case 
that the geometries converge before their total energies do. It is thus sufficient to perform the search at relatively rough
settings with the benefit of decreasing the computational cost. To ensure that correct results are obtained, a number of 
the most stable solutions can then be investigated using more accurate settings. 

The procedure we employ for this Ru$_3$N$_4$C$_4$ search problem is
\begin{itemize}
    \item Run the GOFEE search for a number of independent searches with rough settings. 
    \item Identify the most stable geometries.
    \item Select those with unique graphs.
    \item Locally optimize these, without constraining the template atoms, with the rough settings followed by the fine settings.
    \item Select again those with unique graphs.
    \item Obtain total energies at higher level of theory, including spin polarization.
\end{itemize}
In the steps involving local relaxations with either rough or fine settings after the GOFEE
search, all atoms are included, i.e.\ both template atoms and the
atoms placed in the search.

In the present work, we have performed 25 independent searches for 1000 iterations, resulting in 25000 structures among which 8685 
are within 2 eV of the most stable structure found during the search. These 8685 structures share 120 unique graphs, when 
using the spectral graph technique described in Section \ref{sec:graph_spectral}.
The best structures of each graph is then relaxed in the target potential. Before performing the 
spin-polarized calculations the graph comparison method was employed again, now with the 120 locally optimized structures, 
resulting in a final total of 28 unique structures. Spin polarized calculations were performed with the 
total magnetic moment fixed at 0, 1, 2 and 3 and the initial magnetic moments distributed evenly among the Ru atoms for each of these 28 structures. \\

\begin{figure*}
    \centering
    \includegraphics[width=0.8\textwidth]{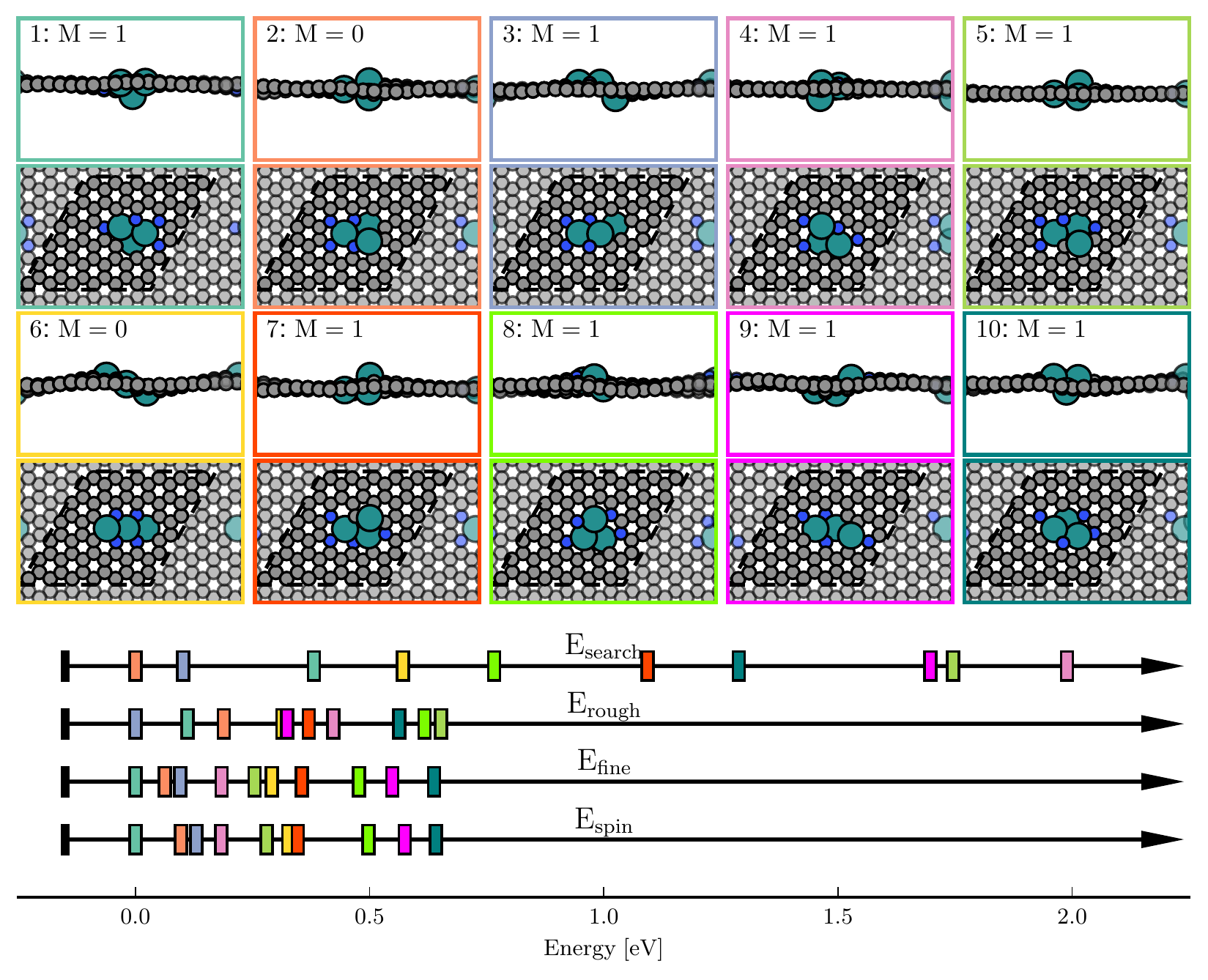}
    \caption{Most stable structures found for search of Ru$_3$N$_4$ in a graphene sheet. The bar plot shows the energy relative
    to the most stable structure at that step of the procedure. The most stable structure identified by the search is structure two, 
    but at the higher level of accuracy structure one has a lower energy. Note that because structure 0 lowers its energy more 
    by the inclusion of spin than the other structures, the relative energy of the other structures including spin is higher than 
    when spin is not included even if their energy also decrease. The total magnetic moment M that leads to the lowest total energy 
    is reported along with each structure.}
    \label{fig:Ru3N4_structures}
\end{figure*}

The 10 most stable structures found by employing this procedure are depicted in Figure \ref{fig:Ru3N4_structures}. 
For the search, the most stable structure found is structure two, however after local optimization with the same settings 
as used for the search, without fixing the graphene sheet, structure three becomes the most stable. Structures four and five both 
decrease their total energy by over 1 eV as a result of the local optimization, going from being uncompetitive structures 
to being possible candidates for the global minimum. This can be attributed to local optimization of the graphene sheet 
being particularly favorable for these two structures. With more accurate DFT settings structure 1 becomes the most stable 
structure and the inclusion of spin further decreases its energy relative to the other structures. Structure one, two, three and five 
all show the triangular arrangement of the Ru atoms that is expected from the experimental results presented by Shufang et. al \cite{Shufang2017}.
We note that in their work, a structural model several eV above the ones shown in Figure \ref{fig:Ru3N4_structures} was proposed. 
This highlights the need for efficient and easy to use global optimization algorithms, as even with experimental evidence as a guide,
guessing the global minimum structure for such complex systems is practically impossible.

\begin{figure} 
    \centering
    \includegraphics[width=0.45\textwidth]{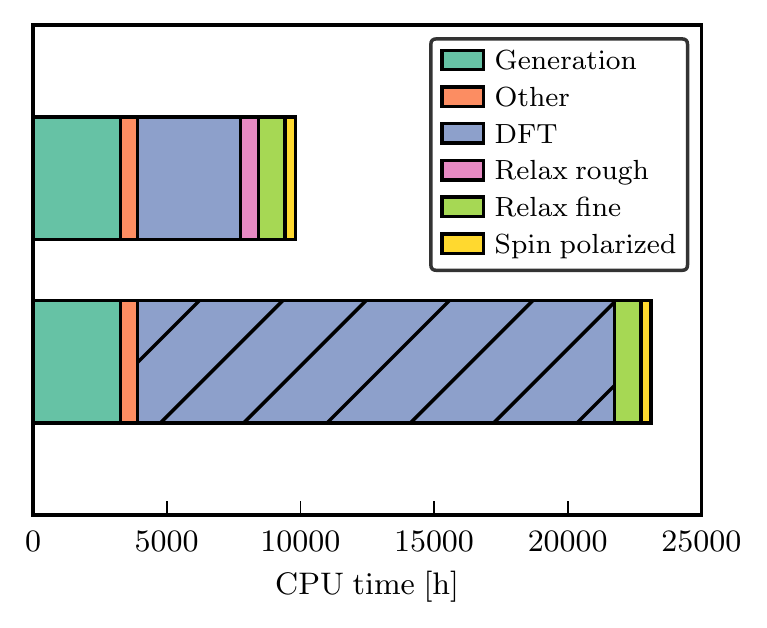}
    \caption{Breakdown of CPU for the searches performed for
      Ru$_3$N$_4$C$_4$-graphene system. The top bar shows the time for the search 
    performed with fast DFT settings, whereas the bottom bar reports the time if it had been performed with the more accurate DFT settings.
    A total of $28 \times 4 = 112$ spin-polarized calculations were performed at roughly 3 CPU hours per calculations, compared to the roughly 
    10 and 40 CPU minutes required for at the low and high unpolarized DFT settings. The hatched area is an estimate based on a single search run with 
    the fine DFT settings.}
    \label{fig:Ru3N4_timing}
\end{figure}

This analysis shows that some degree of reordering of the stability hierarchy of structures must be accounted for when using less accurate settings 
for the potential employed in the search compared to the final desired level of accuracy. 
This begs the question of why not performing the search at the desired level of accuracy? 
An answer to that question is shown in Figure \ref{fig:Ru3N4_timing} which shows that at least twice the computational time budget 
would be required to do so. The post analysis procedure of local optimization and spin-polarized calculations is 
relatively cheap as it is only done for the structures identified by the spectral clustering technique. 

\section{Conclusion}
Global optimization is an essential part of the computational treatment of materials. However, its successful application 
requires choosing an appropriate algorithm according to the difficulty of the problem and the computational demand of the 
chosen potential. Furthermore, computational modeling of materials in general is a rapidly developing field, largely due to the advent of 
machine learning techniques, it is therefore necessary for software tools to enable these developments. 
We have introduced the Atomistic Global Optimization X framework and accompanying Python code for the 
global optimization of atomistic structures that leverage modern programming principles and state of the art machine 
learning techniques to efficiently solve these tasks. The code is flexible and allows for the rapid development and testing of global optimization algorithms.
The application of the package to four examples of global optimization tasks has been documented, one using a simple 
effective medium theory potential that allows for reproduction of the results with a fairly small computational budget and without installation of additional software.
The second example documents how AGOX allows surgically changing
algorithms to reduce the computational demand. We also present an 
application using parallel-tempering that shows how AGOX can take advantage of computational resources through parallelization. 
Finally, an example show-cases the use of AGOX for a real-world atomistic optimization problem. 

\section{Acknowledgements}
This work has been supported by VILLUM FONDEN through Investigator grant, project no. 16562, and by the Danish National Research Foundation through the Center of Excellence “InterCat” (Grant agreement no: DNRF150).
 
\section{Data availability}
Version 1.1.0 of the code is publically available at \agoxrepo \ under a \license \ license. Documentation available at \agoxdocu. Data supporting the findings presented 
in this paper available at \agoxdata. 

\section{Methods}
\label{sec:methods}

\subsection{Gaussian Process Regression}
\label{sec:GPR}
When employing a machined learned model in AGOX we follow Ref.\
\onlinecite{Bisbo2020} and use a gaussian process regression (GPR)
model. With a GPR model, the energy prediction for a structure with
feature $\mathbf{x_*}$ is made as:
\begin{equation*}
    E(\mathbf{x_*}) = K(\mathbf{x_*},\mathbf{X})[K(\mathbf{X},\mathbf{X})+\sigma_n^2\mathbf{I}]^{-1}(\mathbf{E}-\mu(\mathbf{X})) + \mu(\mathbf{x_*})
\end{equation*}
where $\mu$ is the prior, and $\mathbf{E}$ are the energies of training structures that are described by their feature representations $\mathbf{X}$ 
that are compared by the kernel $K$ for which we use a double Gaussian, as in Bisbo et. al. \cite{Bisbo2020}, where an element is calculated as

\begin{align*}
K(\mathbf{x}, \mathbf{x}_*) = \ \theta_0 \bigg [ (1-\beta) &  \exp{\left(-\frac{(\mathbf{x}-\mathbf{x}_*)^2}{2l_1^2}\right)} \\
            + \beta & \exp{\left(-\frac{(\mathbf{x}-\mathbf{x}_*)^2}{2l_2^2}\right)} \bigg ]
\end{align*}

where the length-scales $l_1$ and $l_2$ are chosen such that $l_1 > l_2$ with $\beta = 0.01$. The GPR model also allows the 
calculation of the model uncertainty for a query structure 
\begin{equation*}
    \sigma(x_*) = K(\mathbf{x}_*, \mathbf{x}_*)-K(\mathbf{X}, \mathbf{x}_*)^T[K(\mathbf{X}, \mathbf{X})+\sigma^2_nI]^{-1}K(\mathbf{X}, \mathbf{x}_*),
\end{equation*}
with $\sigma_n$ being a noise parameter.

\begin{figure}
    \centering
    \includegraphics{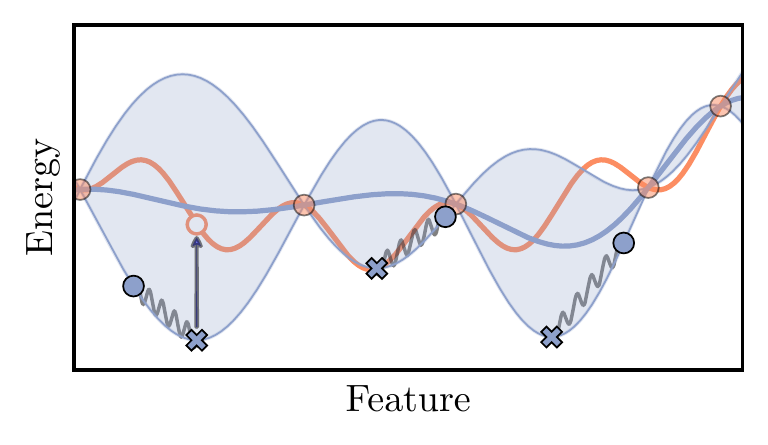}
    \caption{Example of GPR model and LCB sampling. Orange 
    is the real function from which the orange training points have been gathered, 
    this leads to the blue GPR model where the shaded area represents the 
    uncertainty of the model. By sampling a number of points and relaxing in the LCB surface the 
    next point to evaluate and add to training data may be chosen}
    \label{fig:gpr_figure}
\end{figure}

Since the GPR model can estimate the uncertainty, it allows for the
use of the lower-confidence-bound (LCB) acquisition function. This is
done in GOFEE where candidates are relaxed in the LCB and where the next structure to be evaluated in 
the target potential is chosen according to 
\begin{equation}
    x_a = \argmin_{x \in X} \left[ E(x) - \kappa \sigma(x) \right]
\end{equation}
where $X$ is a set of candidate coordinates and $\kappa$ is a parameter decided upon prior to starting the search. 
This procedure is depicted in \ref{fig:gpr_figure}, here three candidate coordinates are generated and locally 
optimized in the LCB expression with the most promising one, that is the one with the lowest LCB value, picked for 
evaluation.

\subsection{Fingerprint feature}
\label{sec:fingerprint}
The representation we employ for the GPR model is the Oganov-Valle fingerprint feature \cite{Valle2010}, where the radial components 
between species A and B are given by
\begin{equation}
    F_{AB}(R) \propto 
    \begin{cases}\displaystyle
      \sum_{ij} \frac{1}{R_{ij}^2} \exp{\left(-\frac{(R-R_{ij})^2}{2\sigma^2}\right)}
      &, R< R_R,\\
      0&, R\ge R_R,
    \end{cases}
  \end{equation}
where $R_R$ (6 \AA) is a hard cut-off and $\sigma$ (0.2 \AA) is a smearing parameter. A feature vector is constructed by sampling the feature 
at intervals of $\Delta$ (0.2 \AA), for multiple species the vectors can are appended together. We also employ angular components, given 
by 
\begin{equation}
    F_{ABC}(\theta) \propto \sum_{ijk} f_c(r_{ij})f_c(r_{ik}) \exp{\left(-\frac{(\theta-\theta_{ijk})^2}{2l_{\sigma}^2}\right)}
\end{equation}
where $l_\sigma$ (0.2 rad) is a smearing parameter and $f_c(r)$ is a cut-off function that ensures that the feature smoothly goes to zero at $R_\theta$ (4 \AA) as controlled by the 
parameter $\gamma$ (2), in particular it is 
\begin{equation}
    f_c(r) = 1 + \gamma (\frac{r}{R_\theta})^{\gamma+1} - (\gamma + 1)(\frac{r}{R_\theta})^\gamma.
\end{equation}
The values given in parenthesis next to each parameter is the value used in this work.

\subsection{Spectral clustering of atomistic structures}
\label{sec:graph_spectral}
Graphs are a natural way of describing atomistic structures, which ball-stick depictions of molecules illustrates clearly. 
For our purposes graphs are particularly useful in order to analyse the vast amount of data generated by search algorithms, 
in order to filter out the unique structures found. We do this by building the adjacency matrix of each structure 
\begin{equation}
    \label{eq:adjacency}
    A_{ij} =
    \begin{cases}   
        0 & \mathrm{if} \ i = j \\        
        1 & \mathrm{if} \ D_{ij} < 1.3 \ d_{cov}(t_i, t_j) \\
        0 & \mathrm{else}
    \end{cases}
\end{equation}
with $t_i$ being the atomic number of atom $i$ and where $D_{ij}$ is the distance between atoms $i$ and $j$, and $d_{cov}(t_i, t_j)$ is the sum of covalent radii for atoms 
of type $t_i$ and $t_j$. From this a Laplacian matrix may be constructed as 
\begin{equation}
    L_{ij} = (\sum_j A_{ij})\delta_{ij} - A_{ij}
\end{equation}
where $\delta_{ij}$ is the Kronecker delta. Finally the eigenvalues of the Laplacian $\mathbf{\lambda} = \lambda_1, \lambda_2, .., \lambda_n$ may be computed and two 
structures may be compared by checking that their eigenvalue spectra are equal. Because this procedure is based 
on the adjacency matrix small variations in bonds lengths do not change the eigenvalues (unless that change causes a 
bond to break according to Eq. \eqref{eq:adjacency}), which is a helpful property when using this feature to find unique 
structures. While the eigenspectrum of distinct graphs can be equal, this is unlikely to be the case in practice \cite{Wilson2008, Wills2020}. 
We use this feature to find distinct structures by grouping together all those structures that have equal eigen spectra. Whereas with a more usual continuous atomistic feature, such as the fingerprint feature used for the 
GPR model, a distance threshold parameter must be chosen. The fingerprint feature describes minute changes in the structure that do 
not correspond to configurational differences, whereas the graph eigen spectrum for structures with variations in bond lengths is equivalent and therefore a distance threshold is not 
necessary. 

\section{References}
\bibliography{bib}

\end{document}